# Impact of Junction Depth and Abruptness on the Activation and the Leakage Current in Germanium n$^+$/p Junctions

William Hsu, Amritesh Rai, Xiaoru Wang, Yun Wang, Taegon Kim, and Sanjay K. Banerjee, *Fellow, IEEE*

*Abstract*— The phosphorous activation in Ge n$^+$/p junctions is compared in terms of junction depth, by using laser spike annealing at 860°C for 400μs. The reverse junction leakage is found to strongly depend on the abruptness of dopant profiles. A shallow and abrupt junction is shown to have lower phosphorous activation level, due to surface dose loss, and higher band-to-band tunneling (BTBT) leakage, when compared to the deep junction. Simulations were carried out to evaluate the lowest achievable OFF-state currents ($I_{OFF}$) for Ge double-gate FETs when using such an abrupt junction. Our results indicate that a Ge body thickness smaller than 5 nm is required to suppress the BTBT leakage and meet the requirement for the high performance devices defined by the International Technology Roadmap for Semiconductors ($I_{OFF}$ = 10$^{-7}$ A/μm).

*Index Terms*—Ge, dopant activation, n$^+$/p junction, laser spike annealing, reverse junction leakage, band-to-band tunneling.

## I. INTRODUCTION

GERMANIUM-channel nMOSFET with high electron mobility and sub-nm EOT high-*k* gate dielectric is a promising candidate for post-Si CMOS [1], [2]. The scalability of Ge devices has been investigated with advanced architectures [3]. However, there are still several remaining challenges. First, achieving highly activated Ge n-type junctions is known be difficult due to the strong Coulomb interactions between substitutional n-type dopants and negatively charged Ge vacancies [4]. Such interactions also enhance dopant diffusion, resulting in dopant loss and deep junction depth. Recently, the use of advanced activation schemes (e.g. laser annealing) has been shown to improve the activation of Ge n-type junctions with good diffusion control [5]–[7], but the performance of Ge shallow junctions is still not satisfactory [8]. A shallow and highly activated junction is essential for highly-scaled devices to reduce short channel effects and the contact resistance between metal and semiconductor.

Another issue is the excessive leakage in Ge junctions, even though Ge can have less implantation-induced extended defects (i.e., end-of-range defects) compared to Si due to its high equilibrium concentration of vacancies [9], [10]. The high intrinsic carrier concentration ($n_i$) of Ge increases generation-recombination and diffusion currents, whereas the small bandgap ($E_g$ = 0.66 eV) gives rise to band-to-band tunneling (BTBT), which can become the dominant leakage mechanism at high electric field (e.g. gate-induced drain leakage, GIDL) [11]. An optimum substrate doping concentration in the range of 1–5×10$^{17}$ cm$^{-3}$ has been suggested [12].

In this work, we compared activation behaviors and leakage currents between the deep and the shallow Ge n$^+$/p junctions, activated by laser spike annealing (LSA). Using the experimental data, the projected lowest achievable OFF-state currents limited by BTBT in Ge n-channel double-gate (DG) MOSFETs were investigated via numerical simulations.

## II. EXPERIMENTS

Experiments were performed on (100) *p*-type Ge wafers, with a background concentration of ~5×10$^{17}$ cm$^{-3}$. For the deep junction, P implant with a dose of 4×10$^{15}$ cm$^{-2}$ and energy of 27 keV was performed, resulting in an amorphous layer thickness of ~64 nm. For the shallow junction, the wafer first received a Ge pre-amorphization implant (PAI) with a dose of 2×10$^{14}$ cm$^{-2}$ and energy of 20 keV, which amorphizes to a depth of ~30 nm, followed by P implant with a dose of 5×10$^{14}$ cm$^{-2}$ and energy of 3 keV. The LSA was performed at the peak temperature of 860 °C for 400μs using a CO$_2$ laser radiation source. The substrate standby temperature is 200 °C. There was no capping layer on the implanted surface during the annealing, and all the samples were fully recrystallized after annealing, as measured by optical ellipsometry.

The chemical concentration of P dopants was studied by secondary ion mass spectrometry (SIMS). The samples for diode and refined transfer length method (RTLM) measurements [13] were prepared by the mesa etch process for a depth of ~500 nm. Then the sidewall was passivated with Al$_2$O$_3$/GeO$_x$ using atomic layer deposition at 200 °C, followed by oxide etch and Ni contact formation.

The device fabrication was supported by the National Science Foundation National Nanotechnology Coordinated Infrastructure (NNCI) Texas Nanofabrication Facility and by Samsung.

W. Hsu, A. Rai, T. Kim and S. K. Banerjee are with the Department of Electrical and Computer Engineering and Microelectronics Research Center, The University of Texas at Austin, Austin, TX 78758, USA (e-mail: william.hsu@utexas.edu).

X. Wang and Y. Wang are with the Ultratech Inc., San Jose, CA 95134 USA (e-mail: ywang@ultratech.com).

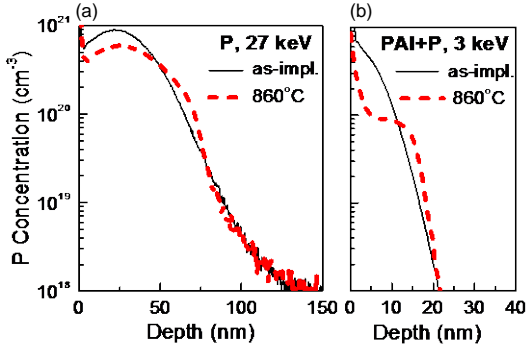

Fig. 1. SIMS profiles of P obtained from samples implanted with (a) P at 27keV (deep junction) and (b) PAI+P at 3keV (shallow junction) before and after laser spike annealing at 860 °C for 400μs.

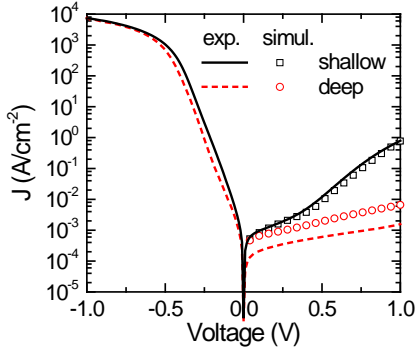

Fig. 2. *I-V* characteristics of the deep (dashed line) and the shallow (solid line) Ge n+/p junctions, activated by laser spike annealing at 860 °C for 400μs. The open symbols represent the results from simulations. The measurements and simulations were performed at 298K.

TABLE I. Parameters for the dynamic nonlocal path BTBT model [14]. *A* and *B* are in the unit of $cm^{-3} \cdot s^{-1}$ and $MV \cdot cm^{-1}$, respectively. $E_{1,\Gamma}$ and $E_{1,L}$ are the eigenenergies of the first sub-band for Γ band and L band, respectively. The quantization direction is [001].

| Ge body thickness | Direct BTBT parameters | Indirect BTBT parameters [15] | $E_{1,\Gamma} - E_{1,L}$ |
|---|---|---|---|
| bulk | | | 0.14 eV |
| 10 nm | $A_{dir}$: $2.2 \times 10^{20}$ | $A_{ind}$: $1.67 \times 10^{15}$ | 016 eV |
| 5 nm | $B_{dir}$: $5.6 \times 10^6$ | $B_{ind}$: $6.55 \times 10^6$ | 0.19 eV |
| 3 nm | | | 0.21 eV |

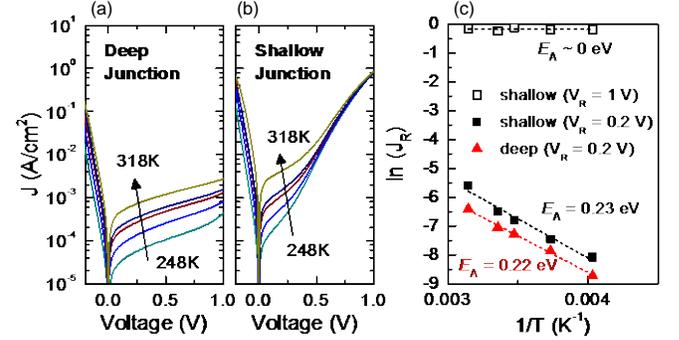

Fig. 3. Temperature-dependent *I-V* characteristics of (a) the deep and (b) the shallow Ge n+/p junctions. The measurement temperatures were 248K, 268K, 288K, 298K, and 318K. Activation energies ($E_A$) were extracted using the (c) Arrhenius plots by the slope of $J_R$ vs. *T* at the reverse bias ($V_R$) of 0.2 V and 1 V.

The simulations of (100) Ge diodes and DGFETs were performed using Synopsys SDevice [14] with physics models including bandgap narrowing, Fermi–Dirac statistics, hydrodynamic transport, doping-dependent mobility, dynamic nonlocal path BTBT and trap-assisted tunneling (TAT), and the Shockley–Read–Hall (SRH) generation/recombination. Quantum correction to the classical carrier densities was carried out using the density-gradient model, with the parameters calibrated from the 1D Schrödinger–Poisson solver [14]. The eigenenergies of the first sub-band of L band and Γ band, obtained from the 1D Schrödinger–Poisson solver, were included in the BTBT model for the calculations of indirect and direct transitions, respectively. For valence band, the quantum correction to the tunneling barriers/potentials was conducted using the density-gradient model. The BTBT parameters for direct transition were extracted from the experimental results in this work, whereas the theoretical parameters from [15] were adopted for indirect transition along <100> direction (**Table I**). The effective tunneling mass of $0.12m_0$ was used for the TAT model. The Ge DGFETs parameters used throughout the simulation are as follows: p-type Ge channel ($1 \times 10^{17}$ cm$^{-3}$); n-type Ge source/drain; 0.9 nm gate EOT; 3.1 eV gate oxide barrier (SiO$_2$); 15 nm channel and gate length; and 0.6 V supply voltage.

## III. RESULTS AND DISCUSSION

The as-implanted and annealed P depth profiles of the deep and the shallow junctions are shown in **Fig. 1(a)** and **(b)**, respectively. After annealing, concentration-enhanced diffusion toward the substrate and dose loss due to P out-diffusion are seen for both the junctions. The change of the junction depth after annealing is very minimal due to the short dwell time of 400μs. However, a much greater dose loss is observed for the shallow junction compared to the deep junction (~57% vs. ~10%).

RTLM measurements were used for the extraction of sheet resistance ($R_s$) at low bias voltage (≤ ±0.5 V) to suppress the substrate leakage, as will be discussed later. Active concentrations ($N_{act}$) were estimated using depth profiles, extracted $R_s$, and a semi-empirical doping-dependent mobility model [16]. The $R_s$ of the deep junction was measured to be 45 Ω/sq, corresponding to an $N_{act}$ of ~ $1.5 \times 10^{20}$ cm$^{-3}$, whereas an $R_s$ of 406 Ω/sq was obtained for the shallow junction, corresponding to an $N_{act}$ of ~ $4.8 \times 10^{19}$ cm$^{-3}$. The $N_{act}$ of the shallow junction is almost the same compared to our previous work using LSA ($N_{act}$ ~ $4.9 \times 10^{19}$ cm$^{-3}$), where the shallow junction was implanted with P alone at 3 keV without PAI [17], and the value still cannot meet the ITRS requirement ($N_{act}$ > $7 \times 10^{19}$ cm$^{-3}$ [18]). These results indicate that the Ge surface plays an important role during activation annealing, strongly altering the density of point defects. High dose loss and low $N_{act}$ in the shallow junction may be due to the vacancy injections from surface. Adding a capping layer has been shown to suppress the dose loss and may help improve the activation of shallow junctions [19].

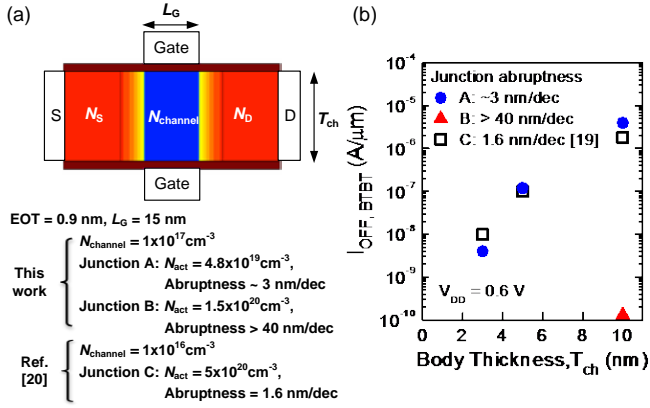

Fig. 4. (a) Schematic of the Ge (100) DGFET structure and parameters used in the simulations ($N_S = N_D = N_{act}$). Junction A and B (n$^+$/p) correspond to the shallow junction and the deep junction in this work, respectively, whereas junction C (p$^+$/n) was assumed in [20]. (b) $I_{OFF,BTBT}$ as a function of body thickness at $V_{DD} = 0.6$V, while considering different junctions (junction A, B and C).

**Fig. 2** shows the diode I–V characteristics measured from the deep and the shallow junctions. Low ideality factors ~1.1 are seen for both the junctions. At the reverse bias of 1 V, the leakage current of the shallow junction is about three orders of magnitude higher compared to the deep junction. To evaluate the carrier conduction mechanism, **Fig. 3(a)** and **(b)** illustrate the temperature dependence of the reverse leakage currents ($J_R$) for the deep and the shallow junctions, respectively. At low reverse bias ($V_R = 0.2$ V), the activation energies ($E_A$) for the deep and the shallow junctions, estimated by the slope of $J_R$ vs. $T$ plots (**Fig. 3(c)**), are about 0.1 eV smaller than half of the band gap of Ge, indicating the mechanism of TAT. However, at high reverse bias ($V_R = 1$ V), the $E_A$ for the shallow junction decreases to ~ 0 eV, and thus the $J_R$ is mainly contributed by the BTBT current. From the diode simulation of the shallow junction (**Fig. 2**), TAT governs the $J_R$ at lower bias (< 0.3 V), and then BTBT becomes dominant with increasing bias. The simulated leakage current of the deep junction was overestimated by four times, which is possibly due to the uncharacterized P channeling tail below the concentration level of $10^{18}$ cm$^{-3}$ in the depth profile (**Fig. 1(a)**), leading to the overestimation of electric field. Compared to the deep junction, the shallow junction suffers from higher BTBT leakage current at the reverse bias of 1 V due to the larger electric field induced by the highly abrupt junction (~3 nm/dec vs. > 40 nm/dec), despite its lower $N_{act}$.

The junction static-power dissipation for the shallow junction remains below 1 W/cm$^2$ for a supply voltage of 1V (**Fig. 3(b)**), and should not be a bottleneck for the development of Ge technology [12]. In the current Si CMOS, the static-power dissipation is dominated by the subthreshold leakage, but it can be overwhelmed by GIDL if using Ge channel due to its higher BTBT rate. Here, the Ge DGFETs (**Fig. 4(a)**) using the junctions in this work as source/drain with different Ge body thicknesses ($T_{ch} = 3$, 5, and 10 nm) were simulated to evaluate the lowest achievable OFF-state current limited by BTBT ($I_{OFF,BTBT}$). The evaluation results are shown in **Fig. 4 (b)** as a function of the $T_{ch}$ against the data from [20], where a similar device with hypothetical source/drain junctions was simulated using a more detailed quantization model. Although the data from [20] are with respect to pFET, the authors of [20] pointed out that their results are not strongly affected by the type of FET (n or p). The leakage currents of the Ge DGFETs are dominated by the direct BTBT for 10-nm and 5-nm body thicknesses, but with 3-nm body thickness, indirect BTBT becomes dominant due to the large energy level difference between the L band and the Γ band (**Table I**). For a 10-nm body thickness, the $I_{OFF,BTBT}$ of the DGFET with the junction abruptness of ~3 nm/dec (junction A) is ~10$^5$ higher than that with the junction abruptness of > 40 nm/dec (junction B), due to the larger electric field at the drain-end of the channel. The $I_{OFF,BTBT}$ can be decreased with $T_{ch}$ due to the increase of the effective bandgap and thus the reduction of BTBT rate. A trade-off between $T_{ch}$ and junction abruptness is demonstrated. To meet the requirement for the high performance devices defined by ITRS ($I_{OFF} = 10^{-7}$ A/μm) [], a Ge body thickness smaller than 5 nm is suggested for a supply voltage of 0.6 V when an abrupt junction is used.

## IV. Conclusions

We showed that although high P dopant activation (> 10$^{20}$ cm$^{-3}$) can be achieved in a deep junction by LSA, the shallow junction exhibited a lower activation level (~4.8×10$^{19}$ cm$^{-3}$) due to the proximity to the surface. The PAI treatment only have a minor impact on the activation of P. In addition, the high electric field in the shallow junction induced by the abrupt dopant profile (~3 nm/dec) leads to a large BTBT-dominated reverse leakage. If such an abrupt junction is used in Ge DGFETs, a Ge body thickness smaller than 5 nm is required to suppress the BTBT leakage for a supply voltage of 0.6 V, as suggested by the numerical simulations.

## Acknowledgment

Helpful discussions with Huang-Sian Lan (National Taiwan University) are gratefully acknowledged.